\documentclass[doublecol]{epl2}

\usepackage{bm}
\usepackage[latin1] {inputenc}
\usepackage{graphicx}
\usepackage{epsfig}
\usepackage{psfrag}
\usepackage[usenames]{color}
\usepackage{color}

\newcommand{\be}{\begin{equation}}
\newcommand{\ee}{\end{equation}}
\newcommand{\ba}{\begin{eqnarray}}
\newcommand{\ea}{\end{eqnarray}}

\begin{document}

\title{Universality class of fiber bundles with strong heterogeneities}
\author{R.\ C.\ Hidalgo\inst{1} \and K.\ Kov\'acs\inst{2,3} \and I.\
Pagonabarraga\inst{4} \and F.\ Kun\inst{2}}
\shortauthor{R.\ C.\ Hidalgo \etal}
\institute{
	\inst{1} AMADE, Departament de F\'{\i}sica \and  Departament de
	         Enginyeria Mec\`anica de la Construcci\'o Industrial \\
	         Universitat de Girona Ave. Montilivi s/n,
	         17071-Girona, Spain \\ 
        \inst{2} Department of Theoretical Physics, University of Debrecen, 
	         P.O.Box: 5, H-4010 Debrecen, Hungary\\
        \inst{3} Department of Appl.\ Math.\ and Prob.\ Theory, University
                 of Debrecen, P.O.Box: 12, H-4010 Debrecen, Hungary\\         
        \inst{4} Departament de F\'{\i}sica Fonamental, Universitat de
                 Barcelona, Carrer Mart\'{\i} i Franqu\'es 1,
                 08028-Barcelona, Spain 
}

        
\abstract{
We study the effect of strong heterogeneities on the fracture of
disordered materials using a fiber bundle model. The bundle is
composed of two subsets of fibers, {\it i.e.} a fraction $0\leq\alpha
\leq 1$ of fibers is unbreakable, while the remaining
$1-\alpha$ fraction is characterized by
a distribution of breaking thresholds.
 Assuming global load sharing, we show analytically that there exists a
critical fraction of the components $\alpha_c$ which separates
two qualitatively different regimes of the system: below $\alpha_c$
the burst size distribution is a power law with the usual exponent
$\tau=5/2$, while above $\alpha_c$ the exponent switches to a lower
value $\tau=9/4$ and a cutoff function occurs with a diverging
characteristic size. Analyzing the macroscopic response of the system
we demonstrate that the transition is conditioned to disorder
distributions where the constitutive curve has a single maximum and an
inflexion point defining a novel universality class of breakdown
phenomena. 
}
\maketitle

\section{Introduction}
\label{intro}

Damage and fracture of materials occurring under various types of
external loads is a very important scientific problem with an enormous
technological impact. During the last two decades the application of
statistical physics has revealed that heterogeneities of materials'
microstructure play a crucial role in fracture processes \cite{alava}.
To capture the effect of disorder, recently several
stochastic fracture models have been proposed such as the fiber bundle
model (FBM)
and lattice models of fuses or springs
\cite{alava,zapperi_nature,klo97,hansen94,kun00,raul_yamir,hidalgo}. Based on
these models, analytic calculations and computer simulations revealed
that macroscopic 
fracture of disordered materials shows interesting analogies with
phase transitions and critical phenomena having several universal
features independent of specific material details \cite{alava,klo97,kun00,moreno_00,raul_yamir,prad2,feri}. 
It has been found
that under a slowly increasing external load macroscopic failure is
preceded by a bursting activity due to the cascading nature of local
breakings \cite{hansen94,klo97}. 
Since the bursts can be recorded experimentally by the
acoustic emission technique, these precursors addressed the
possibility of forecasting the imminent failure event
\cite{pet94,gar97,hansen_imminent_prl,pradhan_predict}. 
The size distribution of bursts was proven to be a power
law with an exponent which is universal for a broad class of disorder
distributions \cite{hansen94,klo97}. Recently, the robustness of the universality class has
been tested by mixing different types of disorder distributions
\cite{divakaran_2}, and by
introducing a gap into the domain of strength values
\cite{divakaran_1}. However, relevant change of the burst size
distribution was only obtained when introducing a finite lower
threshold for the strength disorder. Increasing the threshold strength
a crossover occurs from a power law of exponent $5/2$ 
to another one with a lower exponent $3/2$
\cite{hansen_imminent_prl,hansen_imminent_pre,hansen_lower_cutoff_2005}.
Divakaran and Dutta have studied  
the critical behaviour of a Random Fiber Bundle Model with mixed
uniform distribution of threshold strengths ~\cite{divakaran_1}.  They
have considered two uniform  distributions separated by a gap.  
The  approach developed in this Letter might be interpreted as the {\it
infinite gap} limit of Divakaran's model.

In the present paper we study the effect of strong heterogeneities on
the process of fracture based on a fiber bundle model. We assume that the
system has two components one of which is characterized by a strength
distribution, while the other one is unbreakable. Varying the fraction
of the two components $\alpha$ under global load sharing conditions,
we show analytically that the presence of unbreakable elements has a
substantial effect on the fracture process of the system both on the
micro- and macro-scales. Very interestingly, we find a critical
fraction $\alpha_c$ where a transition
occurs between two qualitatively different regimes: below the critical
point $\alpha < \alpha_c$ the macroscopic constitutive curve has a
single maximum and the burst size distribution is a power law
with the usual mean field exponent $\tau=5/2$. However, above
$\alpha_c$ the macroscopic response becomes monotonous and the burst
exponent switches to a lower value $\tau=9/4$ with a cutoff function.
Based on the analysis of the macroscopic response of the system, we
show that the transition is conditioned to disorder distributions
where the constitutive curve has a single maximum and an inflection
point defining a novel universality class of breakdown phenomena.

\section{Model}
We consider a set of $N$ fibers which are loaded in parallel. Under an
increasing external load $\sigma_o$ the fibers have a linearly elastic
response with a Young modulus $E=1$ fixed for all
the fibers. In order to capture the large variation of disordered
material properties, we assume that the bundle is composed of two
subsets of fibers with strongly different breaking characteristics: 
A fraction $0\leq \alpha \leq 1$ of fibers is {\it strong} in the
sense that they have an infinite load bearing capacity so that they
never break. However, fibers of the remaining $1-\alpha$ fraction 
are {\it weak} and break when the load on them $\sigma$ exceeds a
threshold value $\sigma_{th}^{i}$, $i=1,\ldots , N_w$, where
$N_w=(1-\alpha)N$ is the number of weak fibers. 
The strength disorder of weak fibers is characterized by the
probability density $p(\sigma_{th})$ and distribution 
function $P(\sigma_{th})=\int_0^{\sigma_{th}}p(x)dx$ of the failure
thresholds. After a weak fiber breaks in the bundle, its load has to
be overtaken by the remaining intact ones. For simplicity, we assume
global load sharing (GLS) (also called equal load sharing) which means
that all the intact fibers share the same load $\sigma$, hence, no stress
concentration occurs around failed regions. 
Under these conditions the constitutive equation of the model can be
written as
\begin{equation}
\sigma_o =  (1-\alpha) \left[1-P(\sigma)\right] \sigma   + \alpha \sigma,
\label{const1}
\end{equation}
where  $\sigma_o$ is the external load acting on the sample and
$\sigma$ denotes the load of single fibers which is related to the
strain $\varepsilon$ of the system as $\sigma=E \varepsilon$. The
first term of Eq.\ (\ref{const1}) 
accounts for the load bearing capacity of the surviving fraction
of {\it weak} elements, and the second one represents the stress
carried by the {\it unbreakable} subset of the system. 
In the following calculations it is instructive to consider two
different strength distributions for the weak fibers, namely, a
uniform distribution between 0 and 1 and a Weibull distribution will
be used with the distribution functions $P(\sigma) = \sigma$ and
$P(\sigma)=1-\exp{\left[-\left(\sigma/\lambda\right)^m\right]}$,
respectively.  

\begin{figure}
\psfrag{aa}{\large $(a)$}
\psfrag{bb}{\large $(b)$}
\begin{center}
\epsfig{bbllx=3,bblly=3,bburx=260,bbury=430, file=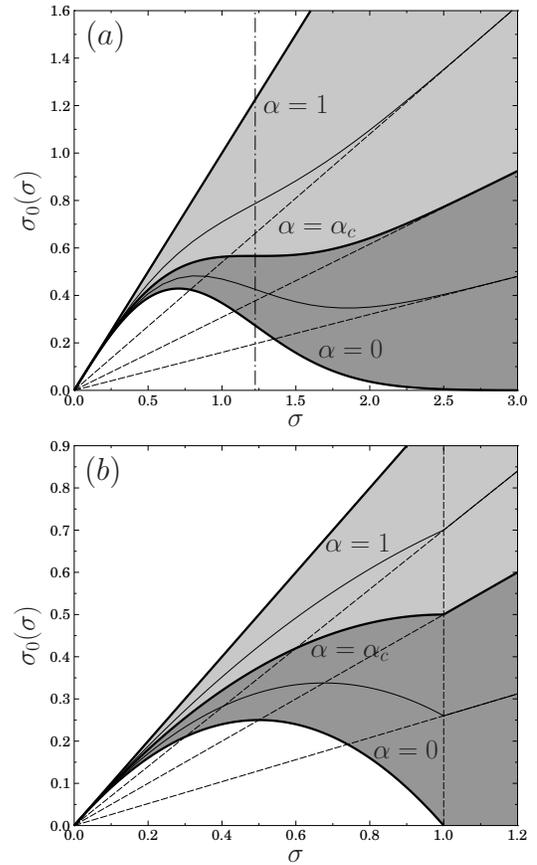, width=7cm}
\caption{Constitutive behavior of the system for several values of
$\alpha$ using a Weibull distribution with $m=2$ and $\lambda=1$ $(a)$,
and uniformly distributed threshold values $(b)$. The two regimes
$\alpha< \alpha_c$ and $\alpha > \alpha_c$ are indicated by the grey
areas, and the dashed lines with slope $\alpha$ show the asymptotic
linear behavior of $\sigma_o(\sigma)$. 
In the figures, the vertical  straight lines represent the position of
the inflexion point and the value of the largest breaking threshold for the Weibull
$(a)$ and uniform distributions $(b)$, respectively.}
\label{fig.1}
\end{center}
\end{figure}
  
The constitutive behavior of the system  is presented in  Fig.\
\ref{fig.1}, for the two different disorder distributions.
We recover the usual FBM solutions \cite{zapperi_nature} in the
limiting case of $\alpha=0$, when the bundle is only composed of weak
fibers. 
Those solutions usually present a parabolic maximum, which  
defines the critical deformation $\sigma_c=E\varepsilon_c$  
and critical strength $\sigma_o^c(\sigma_c)$ of the system. For finite
values of $\alpha$, all the weak fibers break for large enough $\sigma$  so
that the first term of Eq.\ (\ref{const1}) goes to zero
while the {\it unbreakable} fibers overtake the entire external load. 
Consequently, the constitutive curves 
in Fig.\ \ref{fig.1} tend asymptotically to a straight line with slope
$\alpha E$.  It can be   
seen in Fig.\ \ref{fig.1} that for low values of $\alpha$ the local
maximum of $\sigma_o(\sigma)$ prevails but its position
$\sigma_c(\alpha)$ and value  $\sigma_o^c(\alpha)$ are monotonically
increasing with $\alpha$. It is interesting to note that there exists a
well defined critical value of the fraction of the components
$\alpha_c$  above which $\alpha>\alpha_c$ the local maximum disappears
and the constitutive curve becomes a monotonically increasing function
$d\sigma_o/d\sigma>0$. The position of the maximum $\sigma_c$
is obtained from the condition of extreme
$\left.\frac{d\sigma_o}{d\sigma}\right|_{\sigma_c}=0$, which can be cast into
the form 
\begin{equation}
\frac{1}{1-\alpha} = P(\sigma_c) + \sigma_c p(\sigma_c).
\label{eq:alpha_1}
\end{equation}
The above equation should be solved for $\sigma_c$ as a function
of $\alpha$, then $\sigma_o^c$ can be determined by substituting
$\sigma_c(\alpha)$ into Eq.\ (\ref{const1}). 
Since the derivative of the constitutive curve $\sigma_o(\sigma)$ has
a minimum in the inflexion point $\sigma_{in}$, the right hand side of
Eq.\ (\ref{eq:alpha_1}) has a maximum at $\sigma_c=\sigma_{in}$. It
follows that Eq.\ (\ref{eq:alpha_1}) can only be solved for
$\sigma_c(\alpha)$ until $\alpha\leq \alpha_c$, where the critical
fraction of strong fibers $\alpha_c$ is defined as the solution of
$\sigma_c(\alpha_c)=\sigma_{in}$. It is important to emphasize that
the location of the inflexion point $\sigma_{in}$ does not depend on
the value of $\alpha$ since the second derivative of the constitutive
equation $\sigma_o(\sigma)$ reads as
\begin{eqnarray}
\left.\frac{d^2 \sigma_o}{d^2 \sigma }\right|_{\sigma_c} =
-(1-\alpha)\left[2p(\sigma)  + \sigma p'(\sigma)\right].
\label{min3}
\end{eqnarray}
For the case of Weibull distributions the general solution 
$\sigma_c(\alpha)$ cannot be obtained in a closed form. However, one
can still calculate analytically $\alpha_c$ and determine $\sigma_c$ 
for two parameter values $\alpha=0$ and $\alpha=\alpha_c$. The
calculations result in
$\sigma_c(\alpha=0) = \lambda(1/m)^{1/m}$, and $\sigma_c(\alpha_c)=\sigma_{in}$, 
where the inflexion point is $\sigma_{in} = \lambda\left[(1+m)/m\right]^{1/m}$. 
The critical point $\alpha_c$ is then obtained by substituting $\sigma_{in}$ into Eq.\
(\ref{eq:alpha_1}) which yields $\alpha_{c}=m
e^{-(1+m)/m}/\left[1+me^{-(1+m)/m}\right]$.

Note that the above arguments do not apply to the uniform
distribution, since the constitutive curve $\sigma_o =
\left[1-\left(1-\alpha\right)\sigma \right]\sigma$ does
not have an inflexion point  (see Fig.\
\ref{fig.1}$(b)$). 
The position of
the maximum of $\sigma_o(\sigma)$ can be obtained analytically as
$\sigma_{c}(\alpha)=1/2(1-\alpha)$, which holds for  
$\alpha\le\alpha_c$ with the critical value of the control parameter
$\alpha_c=1/2$. At $\alpha_c$ the value of $\sigma_c$ coincides with
the upper bound of strength values $\sigma_{th}^{max}=1$. The
parabolic shape of the constitutive curve prevails even for
$\alpha>\alpha_c$ but $\sigma_o(\sigma)$ becomes linear at
$\sigma=\sigma_{th}^{max}$ before reaching the maximum, so that the
rest of the parabola cannot be realized.  

The presence of the critical point and the qualitatively different
forms of $\sigma_o(\sigma)$ below and above $\alpha_c$  have a
substantial effect on the microscopic breaking of the system. 
Under stress controlled loading conditions the decreasing part of
$\sigma_o(\sigma)$ can not be accessed for $\alpha<\alpha_c$. Contrary,
a horizontal jump occurs giving rise to a large number of breakings in
one step.   
For uniformly distributed failure thresholds, this unstable avalanche is
the last one that includes all the remaining weak fibers. In the
Weibull case, however, the threshold values are distributed over an 
infinite domain, so that the jump is still followed by breaking
events which disappear only asymptotically. For the detailed
characterization of the microscopic breaking process, we analyze the 
size distribution of bursts of fiber breakings. 

\section{Precursory activity}
Under stress controlled loading conditions, each fiber breaking is
followed by the redistribution of load over the intact
elements. Assuming global load sharing the load is
everywhere the same $\sigma$ in the system. When the external load is
increased quasi-statically, {\em i.e.\ } $\sigma_o$ is increased to
break only a single fiber, the subsequent load redistribution triggers
an entire 
burst of breakings. In the simple FBM these local failure events
result in fluctuating burst sizes $\Delta$, with an increasing
average, as macroscopic failure is approached. The size distribution
of the bursts is one of the most important characteristics of the
microscopic fracture process which can be monitored experimentally by
the acoustic emission techniques. 
It has been demonstrated 
that in FBM under GLS conditions \cite{klo97}, the burst
size distribution can be obtained analytically in the form of an
integral
\begin{equation}
\frac{D(\Delta)}{N} =
\frac{\Delta^{\Delta-1}}{\Delta!}\int_0^{\sigma_{m}}
p(\sigma)(1-a_\sigma)a_\sigma^{\Delta-1}e^{-a_\sigma 
\Delta} d\sigma, 
\label{eq:DDelta}
\end{equation}
where $a_\sigma=\sigma p(\sigma)/[1-P(\sigma)]$ is the average number
of fibers which break as a consequence of a single fiber failure at  
the load $\sigma$. It was shown in Refs.\ \cite{klo97,hansen_imminent_prl,hansen_imminent_pre}
that the distribution $D(\Delta)$ simplifies to 
a power law $D(\Delta)\sim \Delta^{-\tau}$ with the exponent
$\tau=5/2$ for a broad class of disorder distributions where the
constitutive curve of the system has a single quadratic maximum.

In the following we show analytically that in the presence of
unbreakable fibers, the avalanche statistics changes and a novel
universality class of FBMs emerges. Slowly increasing the external
load to break a single fiber, its failure stress $\sigma$ is
equally redistributed over the intact fibers giving rise to the load
increment 
\begin{equation}
\delta \sigma=\frac{\sigma}{N\left[1-P(\sigma)\right](1-\alpha)+\alpha
N}. 
\end{equation}
It can be seen that the strong fibers reduce the load
increment $\delta \sigma$ on the weak ones, since the load beared by 
the strong fibers does not contribute to breaking. The average number
of fibers $a_\sigma$ which fail as a consequence of this increment
$\delta \sigma$ can be cast into the form 
\begin{equation}
a_\sigma = (1-\alpha) Np(\sigma)\delta\sigma=\frac{(1-\alpha)\sigma
p(\sigma)}{\alpha+(1-\alpha)\left[1-P(\sigma)\right]}. 
\label{asigma}
\end{equation}
The size distribution of the resulting bursts can be obtained by
substituting Eq.\ (\ref{asigma}) into the general expression Eq.\
(\ref{eq:DDelta}), where we have to analyze the behavior of the
integral 
\begin{equation}
I(\Delta)\equiv\int_0^{\sigma_{m}}p(\sigma)\frac{1-a_\sigma}{a_\sigma}e^{-\Delta
[a_\sigma-\ln a_\sigma]}d\sigma,
\label{eq:Idelta}
\end{equation}
for different values of $\alpha$.  The upper integral 
limit corresponds to the location of the maximum in 
the constitutive curve $\sigma_m$.  For large $\Delta$ this integral is 
controlled by the maximum of the
exponent. The extreme condition of $\psi\equiv a_\sigma-\ln a_\sigma$
result in $\psi ' = a_\sigma' (1-\frac{1}{a_\sigma})=0$,  
corresponding to a maximum  at $a_\sigma=1$. Below the critical point
$\alpha<\alpha_c$, for  $a_\sigma<1$, we can make the expansions 
\begin{equation}
a_\sigma\simeq 1+a'_\sigma \mid_{\sigma_{m}} (\sigma-\sigma_m), 
\label{eq:a1st}
\end{equation}
and
\begin{equation}
\psi \simeq 1+\frac{{a'_\sigma}^2}{2} \mid_{\sigma_{m}} (\sigma-\sigma_m)^2.
\label{eq:psi1st}
\end{equation}
Inserting Eqs.\ (\ref{eq:a1st},\ref{eq:psi1st}) into the expression
of $I(\Delta)$ we get 
\begin{equation}
I(\Delta)\simeq p(\sigma_{m}) a_\sigma'e^{-\Delta}\int_0^{\sigma_{m}}
(\sigma-\sigma_{m}) e^{-\frac{\Delta
a_\sigma'^2}{2}(\sigma-\sigma_{m})^2}d\sigma. 
\end{equation}
Substituting $I(\Delta)$ into the general equation Eq.\
(\ref{eq:DDelta}) and taking the large $\Delta$ limit of the
prefactor, the asymptotic behavior of the burst size distribution can be
cast in the form 
\begin{equation}
\frac{D(\Delta)}{N}\simeq\frac{p(\sigma_m)}{\sqrt{2\pi}a_\sigma'}\Delta^{-5/2},
\label{eq:powold}
\end{equation}
which coincides with the known result of Refs.\ \cite{klo97,hansen_imminent_prl,hansen_imminent_pre} in the limit
$\alpha=0$. This derivation implies that the presence of a finite amount of
unbreakable fibers does not change qualitatively the behavior
of the system while the single quadratic maximum of the constitutive
curve prevails $\alpha<\alpha_{c}$.

The situation drastically changes when we reach $\alpha_{c}$, since at
this point the position of the maximum $\sigma_c(\alpha_c)$ and of the
inflexion point $\sigma_{in}$ of the constitutive curve coincide with
each other so that $\left.\frac{d \sigma_o}{d \sigma}
\right|_{\sigma_{c}} = 0$ and  $\left.\frac{d^2 \sigma_o}{d 
\sigma^2}\right|_{\sigma_{c}} = 0$ hold. Above $\alpha_c$ no maximum
of the constitutive curve exists $d \sigma_o/d \sigma>0$.
It can easily be shown that at $\alpha=\alpha_{c}$ the average number
of failing fibers $a_{\sigma}$ as a consequence of a single fiber
breaking has the properties $a_{\sigma_{c}}=1$ and $a'_\sigma
\mid_{\sigma_{c}}=0$. In order to determine the asymptotic behavior of
$I(\Delta)$ it is then necessary to carry out the Taylor expansions of Eqs.\
(\ref{eq:a1st},\ref{eq:psi1st}) to the next order. In this case we
obtain for $a_{\sigma}$ and $\psi$
\begin{equation}
a_\sigma\simeq 1+\frac{a{''}_\sigma}{2}(\sigma-\sigma_{c})^2, \ \ \
 \ \ \ \psi\simeq 1+\frac{3 a_\sigma{''}^2}{4!}(\sigma-\sigma_{c})^4.
\end{equation}
Inserting these expressions into Eq.\ (\ref{eq:Idelta}), the integral
can be cast into the form
\begin{equation}
I(\Delta)\simeq \frac{p(\sigma_{c})a''_\sigma e^{-\Delta}}{2}\int_0^{\infty}
d\sigma
(\sigma-\sigma_{c})^2e^{\frac{3{a''}_\sigma^2\Delta}{4!}(\sigma-\sigma_{c})^4}.
\end{equation}
Following the usual procedure, we arrive at
\begin{equation}
I(\Delta)\simeq \frac{\Gamma\left(\frac{3}{4}\right)}{2^{5/2}3^{7/4}
{a''}_\sigma^{1/2}}e^{-\Delta}\Delta^{-3/4},
\end{equation}
which implies that at the critical point $\alpha_c$
the  avalanche size distribution changes to
\begin{equation}
\frac{D(\Delta)}{N}\simeq  \frac{\Gamma\left(\frac{3}{4}\right)}{24
\sqrt{3 \pi a''_\sigma}3^{1/4} }\Delta^{-9/4}.
\label{eq:pownew}
\end{equation}
Our derivation demonstrates that increasing $\alpha$ the behavior of
the system changes both on the macro- and the micro-scales. We showed
that while the quadratic maximum of $\sigma_o(\sigma)$ prevails, {\em
i.e.} below the critical point $\alpha_c$, the asymptotic  behavior
of the burst size distribution $D(\Delta)$ is controlled by the
vicinity of the maximum resulting in a power law functional form
$D(\Delta) \sim \Delta^{-\tau}$ with an universal exponent
$\tau=5/2$. 
However, at $\alpha_c$ the constitutive
curve becomes monotonically increasing $d \sigma_o/d \sigma>0$ and the
avalanche statistics is dominated by the inflexion point of
$\sigma_o(\sigma)$, giving rise to a different 
value of the exponent $\tau=9/4$. Varying the
control parameter $\alpha$, the exponent $\tau$ suddenly switches
between the two values $5/2$ 
and $9/4$ when passing the critical point $\alpha_c$. Note that in the
derivation the only assumption we made is that the constitutive curve
of the system has a single maximum and an inflexion point. It follows
that the change of the exponent $\tau$ of the avalanche size
distribution can be observed for a large variety of disorder
distributions defining a novel universality class of breakdown
phenomena. This universality class is narrower than the one in which
the power law behavior of $D(\Delta)$ emerges with the exponent
$\tau=5/2$. For instance, the Weibull distributions do present the
above switching of exponents, however, the uniform distribution does
not.

\begin{figure}
\psfrag{aa}{\large $(a)$}
\psfrag{bb}{\large $(b)$}
\psfrag{cc}{\large $(c)$}
\psfrag{dd}{\large $(d)$}
\begin{center}
\epsfig{bbllx=0,bblly=0,bburx=495,bbury=415, file=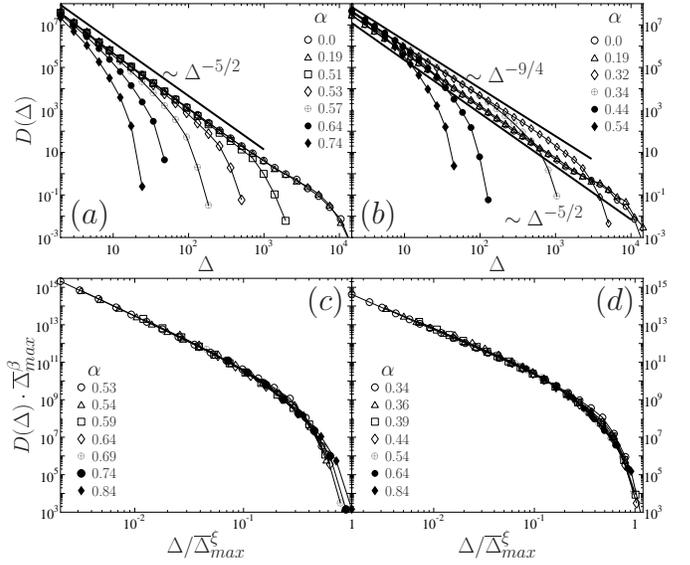, width=8.8cm}
\caption{Non-normalized avalanche size distributions for uniform
$(a,c)$ and Weibull distributions $(b,d)$ varying $\alpha$ below and above
$\alpha_c$. The straight lines in $(a)$ and $(b)$ represent
the power laws obtained analytically Eqs.\
(\ref{eq:powold},\ref{eq:pownew}). Rescaling the two axis according to
the scaling formula Eq.\ (\ref{fun}), a very good quality data collapse
is obtained in $(c)$ and $(d)$.}  
\label{fig.3}
\end{center}
\end{figure}


We have
carried out Monte Carlo simulations to validate the previous
theoretical predictions. We have explored  the quasi-static fracture
process of our fiber  
bundle model using computer simulations of a system composed of
$N=10^6$ fibers and averaging over $10^3$ samples both for uniform and
Weibull  distributions for the fiber breaking thresholds. 
 
\begin{figure}
\psfrag{aaa}{\large $(a)$}
\psfrag{bbb}{\large $(b)$}
\psfrag{ccc}{\large $(c)$}
\psfrag{ddd}{\large $(d)$}
\begin{center}
\epsfig{bbllx=0,bblly=0,bburx=650,bbury=500, file=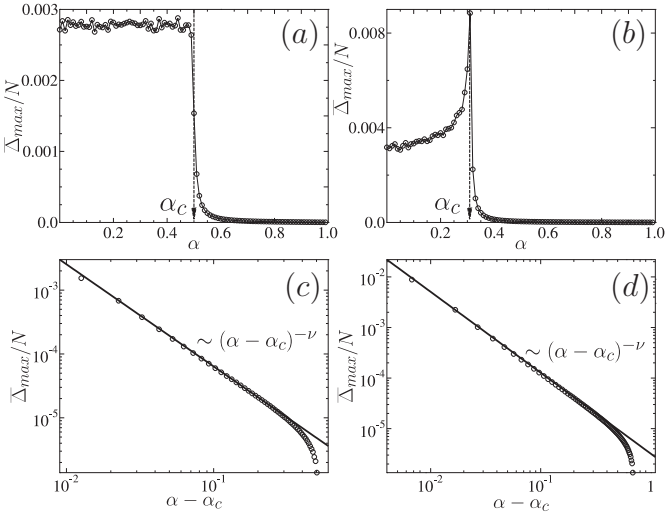, width=8.82cm}
\caption{
The average size of the largest burst ${\overline{\Delta}}_{max}$ as
a function of $\alpha$ for uniform $(a,c)$ 
and Weibull $(b,d)$ distributions. The vertical straight lines in
the figures indicate the corresponding critical point $\alpha_c$.
$(c)$ and $(d)$ show that approaching $\alpha_c$ from above,
${\overline{\Delta}}_{max}$ has a power law divergence as a function
of $\alpha-\alpha_c$. The value of the exponent is $\nu=1.56\pm 0.07$ for
both cases.
}
\label{fig.2}
\end{center}
\end{figure}

Figure \ref{fig.3} displays the burst statistics for a fiber bundle
with uniform and Weibull failure thresholds in Figs. \
\ref{fig.3}($a,c$) and  Figs. \ref{fig.3}($b,d$), respectively. In  both
cases, below the critical fraction of unbreakable fibers, $\alpha_c$,
the burst distributions $D(\Delta)$ do not change significantly, and
even the cutoffs  associated to the lack of numerical statistics  do
not change with $\alpha$. For a uniform distribution of breakable
fibers, the constitutive equation for $\sigma \leq \sigma_c(\alpha)$
reduces to the one corresponding to a material  composed only by  weak
fibers, with threshold values between 
zero and the upper bound $\sigma_{th}^{max} = 1/\left(1-\alpha\right)$.
Hence, it follows that the entire failure process of
the bundle, obtained at different $\alpha$ values, remains the same
until there are enough weak fibers in the system $N_{w}> N/2$. 
Furthermore, for the parameter regime $\alpha< \alpha_c$ the avalanche statistics 
does not change. We obtain the typical power law distribution $D(\Delta)\sim \Delta^{-\tau}$
with the exponent $\tau = 5/2$ (Figs.\ \ref{fig.3}($a$,$c$)).
On the other hand, the parabolic shape of the constitutive behavior 
$\sigma_o(\sigma)$ also prevails for $\alpha > \alpha_c$. 
In this regime, the system behaves as if the loading process was
stopped before reaching the maximum of $\sigma_o(\sigma)$, due to the insufficient 
number of breakable fibers $N_w <N/2$ (compare to Fig.\ \ref{fig.1}$(a)$).  
Consequently, the cutoff of the distribution 
$D(\Delta)$ in Fig.\ \ref{fig.3}$(a)$ decreases with increasing
$\alpha$. However, the exponent $\tau$ keeps the same value as below
$\alpha_c$, in agreement with our predictions and with Ref.\
\cite{klo97}.

We use the average size of the largest burst $\overline{\Delta}_{max}$ as 
the characteristic burst size of the system. It can be seen in Fig.\ \ref{fig.2}$(a)$ 
that for the uniform distribution below $\alpha_c$, the value of 
$\overline{\Delta}_{max}$ is constant, while it decreases rapidly when $\alpha$ surpasses 
$\alpha_c$.  Figure \ref{fig.2}$(c)$ demonstrates that approaching
$\alpha_c$ from above the characteristic burst size shows a power law
divergence ${\bar{\Delta}}_{max} \sim (\alpha-\alpha_c)^{-\nu}$. The value of the 
exponent $\nu = 1.56\pm0.07$ was obtained numerically. 

In Figures \ \ref{fig.3}($b$,$d$), the burst statistics of a fiber bundle with a 
Weibull distribution of breaking thresholds is illustrated. The Weibull parameters 
were set to $m=2$ and $\lambda=1$, corresponding to a critical point $\alpha_c\simeq 0.3085$.
In this case, below the critical point $\alpha<\alpha_c$ the burst size distribution 
has a power law behavior $D(\Delta)\sim \Delta^{-\tau}$, with the exponent
$\tau=5/2$ (Fig.\ \ref{fig.3}$(b)$). In this regime, a slight increase of the 
cutoff burst size appears when increasing $\alpha$, however, the power law part of the 
distribution does not change. When $\alpha$ surpasses $\alpha_c$, 
the exponent of the power law regime of $D(\Delta)$ suddenly 
switches to the lower value $\tau=9/4$. The latter is in excellent 
agreement with our analytic predictions (see Fig.\ \ref{fig.3}$(b)$).  
Moreover, we again find that the characteristic burst 
size $\overline{\Delta}_{max}$ diverges as a power law as we approach 
$\alpha_c$ from above. The value of the critical exponent $\nu$ is the same
as for the uniform case (see Fig.\ \ref{fig.2}$(d)$). 
We emphasize that the value of the exponent of the power 
law regime of $D(\Delta)$ remains constant $\tau=9/4$ when 
changing $\alpha$ above $\alpha_c$. 

Using $\overline{\Delta}_{max}$ as a scaling variable, we introduce
the scaling ansatz
\begin{equation}
D(\Delta) = \bar{\Delta}_{max}^{-\beta} g(\Delta/\overline{
{\Delta}}_{max}^{\xi})
\label{fun}
\end{equation}
for the burst size distributions above the critical point
$\alpha>\alpha_c$. Here $\beta$ and $\xi$ are scaling
exponents, which have the relation $\beta=\tau \xi$ with $\tau=5/2$
and $\tau=9/4$ for the uniform and Weibull distributions,
respectively.  
Figures \ref{fig.3}$(c)$ and $(d)$ present the rescaled 
burst size distributions plotting
$D(\Delta)\overline{\Delta}_{max}^{\beta}$ as a function of 
$\Delta/\overline{\Delta}_{max}^{\xi}$. The high quality data collapse
is obtained with the parameters $\beta=3.25$, $\xi=1.25$ and {\bf $\beta=2.52$,
$\xi=1.12 $,} for the uniform and Weibull distributions, 
which are consistent with the two different values of the $\tau$
exponent. 

\section{Discussion}
Our numerical and analytical calculations revealed that the presence
of unbreakable elements gives rise to a substantial change of the
fracture process of disordered materials both on the micro- and
macro-scales. Astonishingly we found a critical fraction of the
breakable and unbreakable components where the exponent of the burst
size distribution switches from the well known mean field exponent of
FBM $\tau=5/2$ to a significantly lower value $\tau=9/4$. The
 transition is conditioned to disorder distributions where the
macroscopic constitutive response of the system has a single maximum
and an inflexion point, implying a novel universality class of FBM. 
Despite we have considered only unbreakable fibers, our results will hold for 
a finite gap of two threshold distributions down to a certain critical
value of the gap size. Below this critical value our model recovers
the former work of Ref.~\cite{divakaran_1}.
Besides its theoretical importance, the problem has several
implications for experimental studies. New
materials of high mechanical performance are often fabricated by
mixing components with widely different properties. 
For instance, fiber reinforced composites are composed of strong
fibers which are embedded in a carrier matrix.
In this case at the breaking of weak elements, the strong ones act as
the unbreakable component of our model. 

Our detailed analytical and numerical study is restricted to the
quasi-static limit of FBMs where the external load is incremented
in a continuous manner. However, in laboratory experiments only
finite discrete load steps can be realized. Recently, it has been
demonstrated that for large enough load increments the statistics of bursts
changes, i.e.\ under GLS conditions the exponent of the power law
distribution of burst sizes takes a higher value $\tau=3.0$
\cite{pradhan_discrete_1,pradhan_discrete_2}. When the system is a
mixture of weak and 
strong fibers the effect of finite load steps depends on the type of
disorder. For uniformly distributed failure thresholds it is
straightforward to show that the burst exponent changes to $\tau=3.0$
both below and above the critical fraction $\alpha_c$. Nevertheless,
disorder distributions for which the constitutive curve has an
inflexion point (e.g.\ Weibull distribution) deserve a detailed study
which will be presented elsewhere. 

A very interesting application of FBMs is to study the time dependent
deformation and rupture of disordered materials under a constant
external load (creep rupture). To
understand damage enhanced creep 
processes the relaxation dynamics of FBMs has recently been
investigated in details
\cite{kun_creep_1,kun_creep_2,pradhan_relax}. It has been 
found that below the critical load 
$\sigma_c$ the system suffers only partial failure and relaxes to 
a stable state, while above $\sigma_c$ macroscopic breaking occurs in
a finite time. When approaching $\sigma_c$ from either side, the
characteristic time scale (relaxation time and lifetime of the system)
has a power law divergence with a universal exponent $1/2$
\cite{kun_creep_2,pradhan_relax}. Since the 
relaxation dynamics of the system is determined by the functional form
of the constitutive curve in the vicinity of the critical load, novel
behaviour can be expected in the presence of strong fibers. The
relaxation dynamics and creep rupture open up interesting
possibilities for future applications of our model.

\section{Acknowledgments}
This work is part of the Spanish-Hungarian Intergovernmental 
Scientific Project HH2005-0016. RCH acknowledges the financial support  
of the Spanish Minister of Education and Science, through a {\it Ramon y Cajal Program}.
IP thanks Spanish M.E.C. (FIS2005-01299) and DURSI 
({\sl Distinci\'on de la Generalitat de Catalunya}) (Spain)
for financial support. KK and FK were also supported by OTKA T049209
and NKFP-3A/043/04. 

\bibliographystyle{unsrt}

\end{document}